\documentclass[aps,preprint]{revtex4}
\usepackage{amsfonts}
\usepackage{amsmath}
\usepackage{amssymb}
\usepackage{graphicx}
\usepackage{subfigure}
\usepackage{mathrsfs, mathtools}
\usepackage{pdfpages}
\usepackage{multirow}
\usepackage{float}
\usepackage{color}
\usepackage{xcolor}
\usepackage{caption}
\usepackage{lineno,hyperref}
\hypersetup{colorlinks =true, allcolors = blue}
\providecommand{\U}[1]{\protect\rule{.1in}{.1in}}

\begin{document}
\preprint{HEP/123-qed}
\title{ GUP corrected
black holes with cloud of string}
\author{Ahmad Al-Badawi}
\email{ahmadbadawi@ahu.edu.jo 
(corresponding author)}
\affiliation{Department of Physics, Al-Hussein Bin Talal University, P. O. Box: 20, 71111,
Ma'an, Jordan.}

\author{Sanjar Shaymatov}
\email{sanjar@astrin.uz}
\affiliation{Institute for Theoretical Physics and Cosmology, Zhejiang University of Technology, Hangzhou 310023, China}
\affiliation{Institute of Fundamental and Applied Research, National Research University TIIAME, Kori Niyoziy 39, Tashkent 100000, Uzbekistan}
\affiliation{University of Tashkent for Applied Sciences, Str. Gavhar 1, Tashkent 100149, Uzbekistan}
\affiliation{Western Caspian University, Baku AZ1001, Azerbaijan}

\author{Sohan Kumar Jha}
\email{sohan00slg@gmail.com}
\affiliation{Department of Physics, Chandernagore College, Chandernagore, Hooghly, West Bengal, India}

\author{Anisur Rahaman}
\email{manisurn@gmail.com} \affiliation{Department of Physics,
Durgapur Government College, Durgapur, Burdwan 713214, West
Bengal, India}

\begin{abstract}
We investigate shadows, deflection angle, quasinormal modes (QNMs), and sparsity of Hawking radiation of the Schwarzschild string cloud black hole's solution after applying quantum corrections required by the Generalised Uncertainty Principle (GUP).  First, we explore the shadow's behaviour in the presence of a string cloud using three alternative GUP frameworks: linear quadratic GUP (LQGUP), quadratic GUP (QGUP), and linear GUP. We then used the weak field limit approach to determine the effect of the string cloud and GUP parameters on the light deflection angle, with computation based on the Gauss-Bonnet theorem. Next, to compute the quasinormal modes of Schwarzschild string clouds incorporating quantum correction with GUP, we determine the effective potentials generated by perturbing scalar, electromagnetic and fermionic fields, using the sixth-order WKB approach in conjunction with the appropriate numerical analysis. Our investigation indicates that string and linear GUP parameters have distinct and different effects on QNMs. We find that the greybody factor  increases due to the presence of string cloud while the linear GUP parameter shows the opposite. We then examine the radiation spectrum and sparsity in the GUP corrected black hole with the cloud of string framework, which provides additional information about the thermal radiation released by black holes. Finally, our  inquiries reveal that the influence of the string parameter and the quadratic GUP parameter on  various astrophysical observables is comparable, however the impact of the linear GUP parameter is opposite. 

\end{abstract}
\volumeyear{ }
\eid{ }
\date{\today}
\received{}

\maketitle
\tableofcontents
\section{Introduction}

The expected presence of a black hole in the universe was one of
the most peculiar and enigmatic predictions of General Relativity
by Albert Einstein. LIGO collaboration \cite{LIGO1, LIGO2} detected gravitational waves through
physical sensors and confirmed the theoretical predictions of
General Relativity (GR). Additionally, the collaboration saw the first-ever
merging of two black holes \cite{LIGO2}. Capturing of shadow by EHT collaboration \cite{EHT} 
also confirms the presence of the astrophysical black hole. Consequently, astrophysics entered a
new era with these astounding findings of LIGO and EHT collaboration, and research into 
various aspects of black holes gained enormous impetus.

It is not reasonable to regard black
holes as isolated entities. It turns out that realistic black
holes in nature would never be free from interactions with the
surroundings. Thus, in reality, it is in a perturbed state due to
the existence of other fields in the neighborhood. Therefore, studying the 
impact of surrounding matter through perturbation of spacetime metrics is of particular interest. The influence of the quantum
gravity effect on the characteristics of black holes is an
additional significant direction of investigation. Investigation
of both effects through a single framework would be intriguing. In the present endeavor, the quantum effect will be captured
indirectly through the Generalized Uncertainty Principle (GUP) and the surrounding medium considered here is the cloud of string.

An object known as a string is thought to replace the point-like
particles of particle physics in string theory. It is a one-dimensional object. 
A cloud of strings is a distribution or
ensemble of these fundamental strings that may interact and affect
the geometry of spacetime and physical phenomena nearby, according
to several theories \cite{LETE, GHOSH, EHER, LEE, JPM}. The
significance of the string theory for cosmology, quantum gravity,
and the underlying structure of spacetime have all been discussed
concerning this idea. Mathematical formalism like the
Nambu-Goto action, which captures the behavior of the strings in
spacetime and the interactions between the strings and the
surrounding geometry is commonly used in theoretical physics to
describe the dynamics and features of the clouds of strings. With
implications for our knowledge of the early cosmos, black hole
physics, and the unification of fundamental forces, the study of
the cloud of strings \cite{LETE} is an important field of research
in theoretical physics. Even though the idea of a cloud of strings
is theoretical and speculative, it offers an intriguing way to
investigate the essence of reality and possible links between
string theory and empirical occurrences. Letelier \cite{LETE}
pioneered the idea that the source of the gravitational field
might be a cloud of strings, and the clouds of strings aid in our
understanding of the relationships between black holes and string
theory. He provided a precise solution to the Schwarzschild black
hole in General Relativity where a group of strings encircles the
black hole at the same time. Since then, scientists have
increasingly been able to see a cloud of strings 
to be the source of the gravitational interaction at least at the short distances of the order of Planck length,

Black holes release gravitational waves in response to outside
perturbation. Quasi-normal modes (QNMs) are characteristic
frequencies that depend on the details of the geometry and the
type of the perturbation, but not on the initial conditions The
non-vanishing imaginary element of the quasi-normal frequencies
contains information about how black holes relax after enduring perturbation. Black hole perturbation theory \cite{REGGE,
FJZ1, FJZ2, FJZ3, VMON, TEUK} and QNMs become relevant during the
ring-down phase of a black hole merger, the stage where a single
distorted object is formed, and where the geometry of spacetime
undergoes dumped oscillations due to the emission of gravitational
waves. By utilizing gravitational wave astronomy, we now possess
an effective instrument that enables us to test gravity in extreme
conditions, including BH mimickers, gravity alterations, the Kerr
paradigm of General Relativity, and more. The main textbook on the mathematical
features of black holes is Chandrasekhar's monograph
\cite{CHANDRA}, which can be consulted for outstanding reviews on
the subject \cite{KOK, BERT, AZ}. Taking into account the information above, the study of quasi-normal modes for different black holes 
and the effect of quantum correction and the compact of the surrounding medium on it has become an 
interesting subject of investigation. Cai
and Miao  \cite{CAI} computed the quasi-normal modes and
spectroscopy of a Schwarzschild black hole encircled by a cloud of
strings under Rastall gravity. In \cite{NJ} Newman Janis algorithm
has been applied to obtain the solution of the Kerr black hole
surrounded by a cloud of strings in Rastall gravity.
A few recent studies on QNMs for deferment black holes are \cite{QNM1, QNM2, QNM3, QNM4, QNM5}.

We take into consideration a spacetime background developed in
\cite{LETE}, incorporating quantum correction utilizing the GUP \cite{GUP1, GUP2, GUP3, GUP4,
GUP5, GUP6, GUP7, GUP8} to examine the effects of quantum gravity correction on the QNMs. Here, we will look at a generalized algebra with the
addition of a linear and quadratic momentum element to the usual
Heisenberg algebra, which incorporates the minimum measurable
length  and maximum measurable momentum which is acceptable 
to potentially capture the
quantum gravity effect in an indirect manner \cite{LQG1, LQG2,
LQG3, LQG4, LQG5,LQG6}. Here it is developed in such a manner that
it became compatible with the Double Special Relativity (DSR)\cite{DSR1,
DSR2, DSR3, DSR4}. Since there is currently no developed quantum
theory of gravity, this approach is helpful and widely used. Additionally, the
cloud of string surrounding the black hole is something we are
thinking about. Therefore, the use of GUP will be able to capture
a key role that quantum gravity is expected to play. A few recent
application of Linear quadratic GUP to capture quantum gravity
effects in different perspectives are  perused in the papers
\cite{US1, US2, US3, UMA, SUNAN, JOHN}.
Along with the study of the quasi-normal mode, we study the effect of quantum correction on the shadow and sparsity of Hawking
radiation of Schwarzschild black holes surrounded by the cloud of string.

The rest of the manuscript is organized as follows.  In Sect. II, we review the Schwarzschild black hole in the string cloud context and introduce the GUP-corrected black hole. The linear-quadratic GUP modified shadow is investigated in Sect. III.  Next, we consider the deflection angle and compute it around the GUP-corrected
black hole with the cloud of string using the Gauss-Bonnet theorem (GBT) in Sect. IV. The QNMs and the evolution of perturbations are discussed in sections V and VI respectively. Section VII is devoted to the study of sparsity of Hawking emission including greybody factor and Hawking spectrum and sparsity. We end our article with conclusions and discussion in Sect. VIII.

\section{General description of GUP Corrected Black Hole solutions with cloud of string} The metric of the Schwarzschild black hole in the string cloud context is given by \cite{qc1}
\begin{equation}
\text{d}s^{2}=g(r)\text{d}t^{2}-\frac{1}{g(r)} \text{d}
r^{2}-r^{2}(\text{d}\theta ^{2}+\sin^{2}\theta \text{d}\varphi
^{2}) , \label{equ1}
\end{equation}
where
\begin{equation}
g(r)=1-a-\frac{2M}{r},
\label{equ2}
\end{equation}
where $M$ represents the black hole mass, and $a$ is a parameter describing the string cloud.
The prediction of the presence of a minimal length based on several Quantum Gravity techniques such as black hole physics and string theory has numerous fundamental physical implications at very high energies. At the Planck scale, the Schwarzschild radius of a black hole is comparable to the Compton wavelength. 
Recently, a quantum correction to many black hole solutions, including the Schwarzschild one, has been introduced via GUP to avoid singularities in such solutions by imposing a minimum length other than zero \cite{new1}. This correction modifies a black hole's metric, resulting in a matching new horizon. For example, in the case of a Schwarzschild black hole with the cloud of strings, the original Schwarzschild horizon radius $r_h=\frac{2M}{1-a}$ must be substituted by the GUP-corrected radius $r_{hGUP}$. The essential stages for incorporating GUP correction into the black hole metric (\ref{equ1}) are as follows.\\ Consider  a minimum length requirement exists for the GUP, which is defined as: \cite{qc121,qc122,qc123,qc124}
\begin{equation}
\Delta x \Delta p \ge \frac{1}{2}(1-2\alpha \Delta p + \beta (\Delta p)^2),\label{eq3}
\end{equation}
where $\alpha$ and $\beta$ are dimensionless positive parameters. We can obtain a bound for the massless particles as \cite{qc2} 
\begin{equation}
E \Delta x \ge \frac{1}{2},
\end{equation}
then, Eq. (\ref{eq3}) changes to 
\begin{equation}
\varepsilon \ge E \left[ 1-\frac{\alpha}{2\Delta x} + \frac{\beta}{2(\Delta x)^2} + .... \right],\label{eq4}
\end{equation}
where $\varepsilon$ represents the GUP corrected energy. Assuming 
\begin{equation}
\Delta p \sim E \sim M \hspace{1cm} \text{and}\hspace{1cm} \Delta x \sim r_h=\frac{2M}{1-a}. 
\end{equation}
This Eq. (\ref{eq4}), when expressed in terms of mass, looks like this:
\begin{equation}\label{Eq:Mgup}
M_{GUP} \ge M \left(1-\frac{\alpha}{2} \frac{(1-a)}{2M} + \frac{\beta}{2}\frac{(1-a)^2}{(2M)^2}     \right).
\end{equation}
Therefore the GUP corrected event horizon is given by
\begin{equation}
    r_{hGUP}=\frac{2M_{GUP}}{1-a} \ge r_{h} \left(1-\frac{\alpha}{2r_h} + \frac{\beta }{2r_h^2}\right).
\end{equation}
Lastly, from the metric (\ref{equ1}) we can obtain the metric for a GUP-corrected black hole with the cloud of string by replacing the mass $M$ with the corrected GUP mass $M_{GUP}$ namely
\begin{equation}
\text{d}s^{2}=A(r)\text{d}t^{2} -\frac{1}{A(r)} \text{d}
r^{2}-r^{2}(\text{d}\theta ^{2}+\sin^{2}\theta \text{d}\varphi
^{2}) , \label{equ9}
\end{equation}
where the modified corrected GUP metric function 
\begin{equation}
    A(r)=1-a-\frac{2M_{GUP}}{r}. \label{lapse2}
\end{equation}
The GUP corrected horizon radius is given by 
\begin{equation}
r_{hGUP}=\frac{2M_{GUP}}{1-a}=r_h \left( 1-\frac{\alpha (1-a)}{4M} + \frac{\beta (1-a)^2}{8M^2}   \right).
\label{equ11}
\end{equation}
\begin{figure}
    \centering
    \includegraphics{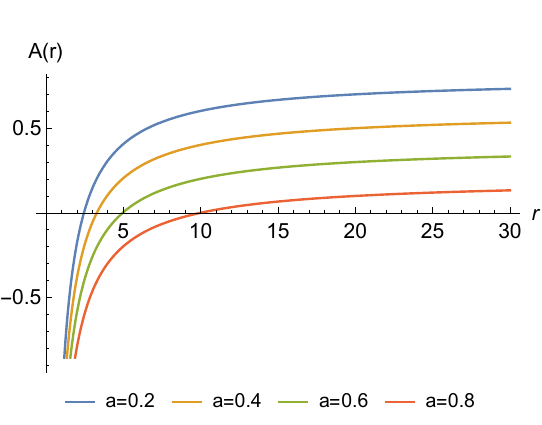}
    \caption{Behaviour of the corrected modified metric function as a function of $r$ for different values of the $a$. ($\alpha=0.1, \beta=0.01$ and $M=1$).}
    \label{figlapse}
\end{figure} Figure \ref{figlapse} depicts the corrected modified metric function's behavior as a function of $r$ for a variety of cloud of strings parameter values. The Fig. shows that the black hole has only one event horizon for various parameter values. There is no other horizon found for the black hole. The graphic indicates that raising $a$ values results in a larger horizon radius.

\section{Linear-Quadratic GUP modified Shadow}
The photon sphere that surrounds a black hole specifies its
shadow, which is determined by certain theoretical parameters.
When we consider spherically symmetric spacetimes, we typically
see a spherical shadow, in contrast to the frequently seen distorted
shadow in the case of rotating black holes. Here, the goal is to investigate the behavior of the shadow in the presence of the
cloud of string within the  GUP framework. The geodesic equations of
a photon traveling in this modified black hole spacetime are constructed as follows.

The generalized Lagrangian is given by the mathematical expression
\begin{equation}
\mathcal{L}(x,\dot{x})=\frac{1}{2}\,g_{\mu\nu}\dot{x}^{\mu}\dot{x}^{\nu}.
\end{equation}
For  a spherically symmetric and static spacetime metric, it can written down as
\begin{equation}
\mathcal{L}(x,\dot{x})=\frac{1}{2}\left[-A(r)\dot{t}^{2}
+\frac{1}{A(r)}\dot{r}^{2}+r^{2}\left(\dot{\theta}^{2}+\sin^{2}\theta\dot{\phi}^{2}\right)\right],
\label{LAG}
\end{equation}
 and the Hamiltonian  corresponding to the Lagrangian
 on the equatorial plane, i.e. when $\theta=\pi/2$ is
\begin{equation}
\mathcal{H}=\frac{1}{2}\left[\frac{p_{t}^{2}}{A(r)}+
A(r)p_{r}^{2}+ \frac{p_\phi^2}{r^{2}}\right] =0, \label{HAM}
\end{equation}
where the over dot denotes the derivative with respect to the proper time $\tau$ and for the black hole spacetime metric
considered here the lapse function is given in (\ref{lapse2}).
Let us now bring  the Euler-Lagrange equation
$\frac{\text{d}}{\text{d}\tau}\!\left(\frac{\partial\mathcal{L}}{\partial\dot{x}^{\mu}}\right)-\frac{\partial\mathcal{L}}{\partial
x^{\mu}}=0$ into action to derive the conserved quantities of the
system, viz. energy and angular momentum. Without any loss of
generality, we can choose the equatorial plane, i.e.
$\theta=\pi/2$. Two killing vectors $\partial/\partial \tau$ and
$\partial/\partial \phi$ render two  conserved  quantity energy
$\mathcal{E}$ and angular momentum $L$ respectively.
\begin{equation}
-p_t=\mathcal{E}=A(r)\dot{t},~~~
p_{\phi}=L=r^{2}\dot{\phi},
\end{equation}
and the radial momentum is given by
\begin{equation} p_r=\frac{\dot{r}}{A(r)}
\label{RADM}
\end{equation}
 The geodesic equation for photon reads
\begin{equation}
-A(r)\dot{t}^{2}+\frac{1}{A(r)}\dot{r}^{2}+r^{2}\dot{\phi}^{2} =
0.\label{GEOE}
\end{equation}
Utilizing the  above equations, we can define
the effective potential of a photon using the radial equation of motion:
\begin{equation}
V_{E}+\dot{r}^{2}=0,
\end{equation}
where the effective potential reads
\begin{equation}\label{e1}
V_{E}=A(r)\left[\frac{L^{2}}{r^{2}}-\frac{{\mathcal{E}^{2}}}{A(r)}\right].
\end{equation}
The unstable circular orbits are specified by the maxima of the
effective potential. Thus these are found in the requirement
 $V_{E}=V^{\prime}_{E}=0$. Using the
condition for the turning point $r_{tp}$ of a photon, where
$\dot{r}=0$ or $V_{E}=0$, the impact parameter $b$ is obtained:
\begin{equation}
b=\frac{L}{\mathcal{E}}=\frac{r}{\sqrt{A(r)}},
\label{IP}
\end{equation}
and using $V_{E}=V^{\prime}_{E}=0$,  we have
\begin{eqnarray}
\frac{A^{\prime}(r)}{A(r)}=\frac{2}{r}\label{RPS}
\end{eqnarray}
where its solution gives us the radius of the photon sphere
\begin{equation}
r_p=\frac{3}{1-a}M_{GUP}.
\end{equation}
Using equation (\ref{IP}) we find that the impact parameter in this
situation is
\begin{equation}
b_g=\frac{9}{2(1-a)}M_{GUP}.
 \label{IPSG}
\end{equation}
It has been found that because of the  string cloud  both $r_p$
and $b_g$ gets scaled by a factor of $(1-a)$. We can also conclude that to get
finite and positive valued shadow radius $1-a$ must be greater than zero, i.e. $a<1$ must be
hold.  On the other hand, using the Eqn. (\ref{GEOE}) within the
orbit equation
\begin{eqnarray}
\frac{\text{d}r}{\text{d}\phi}=\frac{\dot{r}}{\dot{\phi}}=\frac{r^{2}A(r)p_{\phi}}{p(r)},
\end{eqnarray}
we land on the relation
\begin{equation}
\frac{\text{d}r}{\text{d}\phi}=\pm
r\sqrt{A(r)\left[\frac{r^{2}\mathcal{E}^{2}}{A(r) L^{2}}-1\right]}.
\end{equation}
and the above equation reduces to this form
\begin{equation}\label{eosf}
\frac{\text{d}r}{\text{d}\phi}=\pm
r\sqrt{A(r)\left[\frac{r^{2}A(r_{tp})}{A(r)r_{tp}^{2}}-1\right]},
\end{equation}
by using the photon orbit at the turning point
\begin{equation}
\left.\frac{\text{d}r}{\text{d}\phi}\right|_{r=r_{tp}}=0,
\end{equation}
and
\begin{equation}
\frac{\mathcal{E}^{2}}{L^{2}}=\frac{A(r_{tp})}{r_{tp}^{2}}.
\end{equation}
If it is  assumed that light rays are coming from a static
observer located at position $r_{0}$ and transmitting into the
past making an angle $\theta$ with respect to the radial direction then we can write
\begin{equation}
\cot\theta=\frac{\sqrt{g_{rr}}}{\sqrt{g_{\phi\phi}}}\cdot\frac{\text{d}r}{\text{d}\phi}{\Big{|}}_{r=r_{0}}
=\frac{1}{r\sqrt{A(r)}}\cdot\frac{dr}{d\phi}{\Big{|}}_{r=r_{0}}.
\end{equation}
The above equation reduces to
\begin{equation}
\cot^{2}\theta=\frac{r_{0}^{2}A(r_{tp})}{A(r_{0})r_{tp}^{2}}-1,
\end{equation}
which in terms of $sin$ $\theta$ can be rewritten as
\begin{equation}
\sin^{2}\theta=\frac{A(r_{0})r_{tp}^{2}}{r_{0}^{2}A(r_{tp})}.
\end{equation}
The numerical values of $r_p$ and $b_g$ for various values of $\alpha$, $\beta$, and $a$ will now be computed. 
The selection of $\alpha$ and $\beta$ values allows for the demonstration of LQGUP, QGUP, and LGUP.
It is equivalent to linear quadratic GUP (LQGUP) for nonzero values of $\alpha$ and $\beta$, quadratic GUP (QGUP) 
for vanishing $\alpha$, and linear GUP (LGUP) for vanishing $\beta$.
\begin{table} [H]
 \label{newt1} 
 \begin{center} 
\begin{tabular}{ |c|c c|c c|c c| }
 \hline
&   \multicolumn{2}{|c|} {LQGUP:~$\alpha=0.1,\beta=0.01$}  
 &  \multicolumn{2}{|c|}
{QGUP:~$\alpha=0,\beta=0.01$} &    \multicolumn{2}{|c|} {LGUP:~$\alpha=0.1,\beta=0$}\\ \hline
 $a$ & $r_p$ & $b_g$ & $r_p$ & $b_g$ & $r_p$ & $b_g$     \\  \hline 
0.0 & 2.8650 & 4.2975&3.0150 & 4.5225&2.8500 & 4.2750 \\  \hline 
0.2 & 3.6120 &  5.4180& 3.7620&  5.6420& 3.600 &  5.4000 \\  \hline  
0.4 & 4.8590 &  7.2885 & 5.0090 &  7.5135&4.8500 &  7.2750 \\  \hline  
0.6 & 7.3560 & 11.0340& 7.5060 & 11.2590&7.3500 & 11.0250\\  \hline  
0.8 & 11.3580 & 22.2795& 15.0030 & 22.5045&14.850 & 22.2750 \\  \hline 
\end{tabular} 
\caption{Impact of different GUP and the cloud of string on $r_p$ and $b_g$}  \label{tab11}
\end{center}  
 \end{table} 
The data in Table \ref{tab11}  indicate that both $r_p$ and $b_g$ increase with the increase in the value of $a$ for all the GUP (LQGUP, QGUP, LGUP) corrections.

 \section{Deflection angle of GUP corrected black hole with the cloud of string} 
 
In this section we consider the deflection angle and compute it around the GUP-corrected black hole with the cloud of string. For that, we utilize the "Gauss-Bonnet theorem (GBT)" as a new thought experiment addressing the evaluation of the deflection angle for spherically symmetric spacetimes \cite{Gibbons08CQG, Werner12GBT}.   
Here, we note that we consider the weak field approximation when evaluating the deflection angle around the GUP-corrected black hole with the cloud of string. It is to be emphasized that a large amount of work has since been devoted to the study of
the weak deflection angle in various gravity models using the GBT thought experiment  \cite{Ovgun:2018ran,Mandal:2023,Li20,Jusufi18GBT,Zhang21GBT,DCarvalho21,Mustafa22CPC}. Following the GBT method we also further compute the weak deflection angle when restricting null geodesics for the photon motion within the context of the optical metric for the spacetime of the black hole considered here, which is given by 
\begin{eqnarray}
\mathrm{d}\sigma^{2} = g_{kl}^{\mathrm {opt}}\mathrm{d}x^{k} \mathrm{d}x ^{l}
= \mathrm{d}r_{*}^2+\mathcal{A}^2(r_{*})\mathrm{d}\phi^2\, ,
\end{eqnarray}
where 
\begin{equation} \label{Eq:tor1}
\mathcal{A}\big(r_{*}(r)\big)=\frac{r}{\sqrt{A(r)}},
\end{equation}
with $A(r)=\left(1-a-{2M_{GUP}}/{r} \right)$
and the tortoise coordinate $r_{*}$ which reads as 
\begin{equation}\label{Eq:tor2}
r_{*}=\int{\frac{\mathrm{d}r}{1-a-\frac{2M_{GUP}}{r}}}\, ,
\end{equation}
with the corrected mass $M_{GUP}$ given by Eq.~(\ref{Eq:Mgup}). 
Using the aforementioned optical metric, one can further determine the Gaussian curvature $\cal{K}$. To this end it is first valuable to compute the nonvanishing components of Christoffel symbols pertaining to the spacetime metric of black hole \cite{Wald:1984}, which are computed as  
\begin{eqnarray} \label{Eq:Chris1}
\Gamma_{\phi \phi}^{r_{*}}&=&-\mathcal{A}(r_{*})\frac{\mathrm{d}\mathcal{A}(r_{\star})}{\mathrm{d}{r_{\star}}},\\
\Gamma_{r_{*}\phi}^{\phi}&=&\frac{1}{\mathcal{A}(r_{*})}\frac{\mathrm{d}\mathcal{A}(r_{\star})}{\mathrm{d}{r_{\star}}}, \label{Eq:Chris2}
\end{eqnarray}
with the determinant $\det\tilde{g}_{kl}=\mathcal{A}^{2}(r^{\star})$. Taking all together the Gaussian curvature $\mathcal{K}$ can be defined by \cite{Mandal:2023} 
\begin{eqnarray}
\mathcal{K}=-\frac{R_{r_{*}\phi r_{*}\phi}}{\det\tilde{g}_{r\phi}}=-\frac{1}{\mathcal{A}(r_{\star})}\frac{\mathrm{d}^{2}\mathcal{A}(r_{\star})}{\mathrm{d}{r_{\star}}^{2}}\, .
\end{eqnarray}

It must also be noted that the Gaussian curvature $\mathcal{K}$ can be found alternatively within the context of the variable $r$ \cite{Gibbons08CQG,Chandrasekhar:1985,Pourhassan:2022,Sakalli:2016}, i.e., it consequently reads as 
\begin{eqnarray}\label{Eq:GK}
\mathcal{K} & = &-\frac{1}{\mathcal{A}(r^{\star})}\left[\frac{\mathrm{d}r}{\mathrm{d}r^{\star}}\frac{\mathrm{d}}{\mathrm{d}r}\left(\frac{\mathrm{d}r}{\mathrm{d}r^{\star}}\right)\frac{\mathrm{d}A(r)}{\mathrm{d}r}+\left(\frac{\mathrm{d}r}{\mathrm{d}r^{\star}}\right)^{2}\frac{\mathrm{d}^{2}A(r)}{\mathrm{d}r^{2}}\right]\, .
\end{eqnarray}
Here, recalling Eq.~\eqref{Eq:GK} and employing Eqs.~\eqref{Eq:tor1} and \eqref{Eq:tor2} we derive the Gaussian curvature with the equatorial plane area $\mathrm{d}S=\sqrt{|\det{g}_{kl}^{opt}|}\mathrm{d}r \mathrm{d}\phi$ in the optical metric as follows:  
\begin{equation*}
\mathcal{K}\mathrm{d}S=
-\frac{\Big(8 M^2+2 \,\alpha  (a-1) M+\beta(a-1)^2\Big)}{64 (1-a)^{3/2} M^2 r^3} \times
\end{equation*} 
\begin{equation}
\Big(24 M^2-2 (a-1) M (8 r-3 \alpha )+3 (a-1)^2 \beta \Big) \mathrm{d}r\mathrm{d}\phi\, .
\end{equation}
Taking the contribution of the GUP corrected black hole spacetime to the Gaussian curvature, we write the geodesic curvature as  \cite{Gibbons08CQG,Crisnejo18BGT}
\begin{eqnarray}
\frac{\mathrm{d}\sigma}{\mathrm{d}\phi}\bigg|_{C_{R}}=
\left( \frac {r^2}{f(R)} \right)^{1/2}\, .
\end{eqnarray}
Here, we note that $C_{R}$ refers to a curve defined by $r(\phi) = R = constant$ with its radius $R$. To determine the deflection angle using the GBT we further consider the limiting case ${R} \rightarrow \infty$ that results in having the relation $\theta_{0}+\theta_{S}=\pi$ between the object and source.  As a consequence, the above equation in the limiting case can be rewritten as follows:  
\begin{eqnarray}
\lim_{R\to\infty} \kappa_g\frac{\mathrm{d}\sigma}{\mathrm{d}\phi}\bigg|_{C_R}\approx 1\, .
\end{eqnarray}
Taking spatial infinity into consideration $R\to\infty$ and imposing the straight
light approximation $r=b/\sin\phi$, the equation for determining the small deflection angle $\tilde{\delta}$ using the GBT method reads as follows \cite{Gibbons08CQG}:
\begin{eqnarray}
 \int^{\pi+\tilde{\delta}}_0 \left[\kappa_g\frac{\mathrm{d}\sigma}{\mathrm{d}\phi}\right]\bigg|_{C_R}\mathrm{d}\phi-\pi 
 =-\lim_{R\to\infty}\int^\pi_0\int^{\infty}_{\frac{b}{\sin\phi}}\mathcal{K}\,\mathrm{d}S\, .
\end{eqnarray}
It should be noted that we have defined $b$ as the impact parameter in the equation of the deflection angle. Consequently, the deflection angle around the GUP-corrected the black hole with the string cloud background field in the weak
limit approximation can be computed explicitly by the following approximate form
\begin{eqnarray}\label{Eq:def_angle}
\tilde{\delta} \approx
\frac{4 M}{b}+\frac{2 a M}{b}-\frac{\alpha }{b}+\frac{\beta }{2 b M}+\frac{\alpha  a}{2 b}-\frac{3 a \beta }{4 b M}+\frac{\alpha  a^2}{8 b}+\frac{3 a^2 M}{2 b}+\frac{3 a^2 \beta }{16 b M}\, .
\end{eqnarray}
\begin{figure*}
    \centering
{\includegraphics[width=7.75cm]{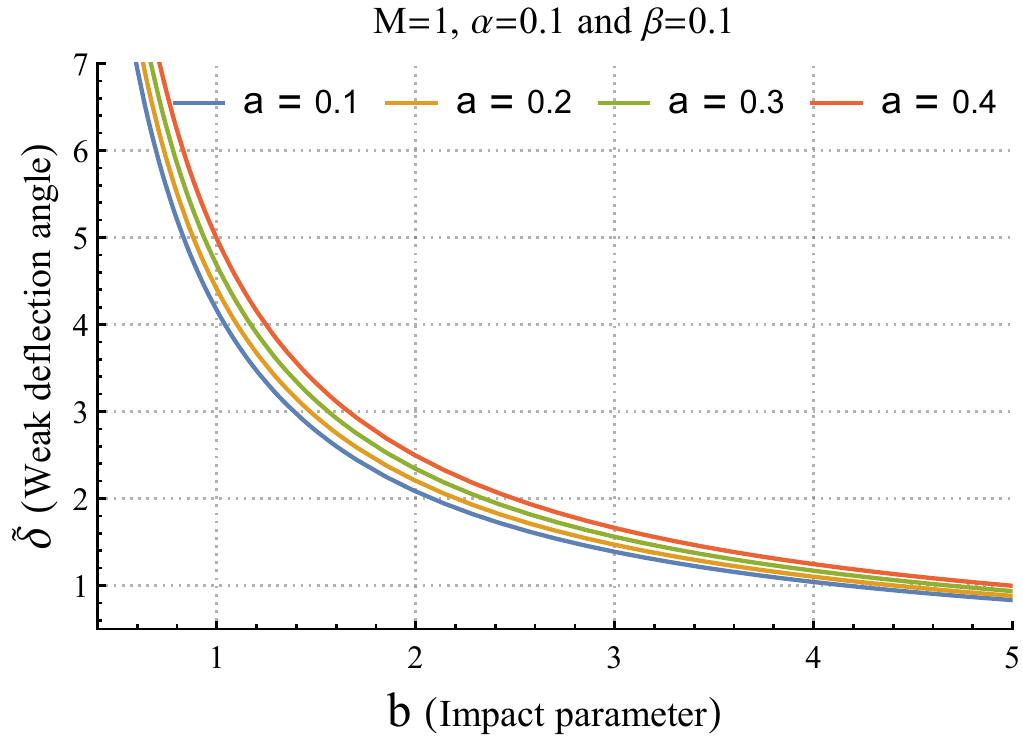} }\qquad
    {{\includegraphics[width=7.75cm]{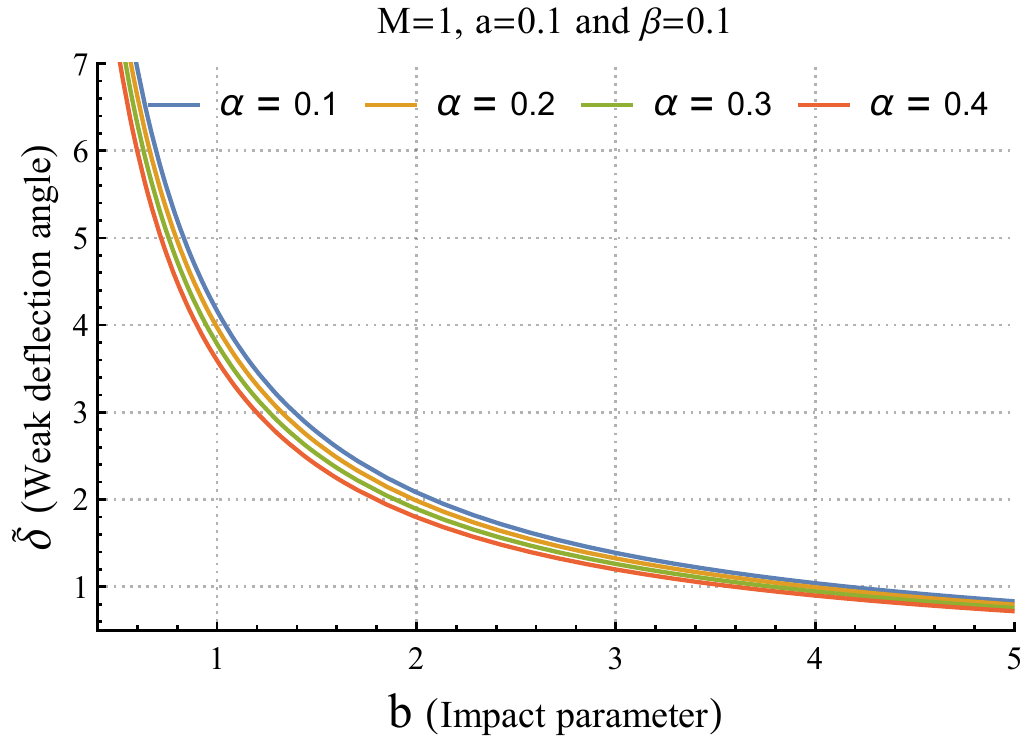}}}\qquad

    {{\includegraphics[width=7.75cm]{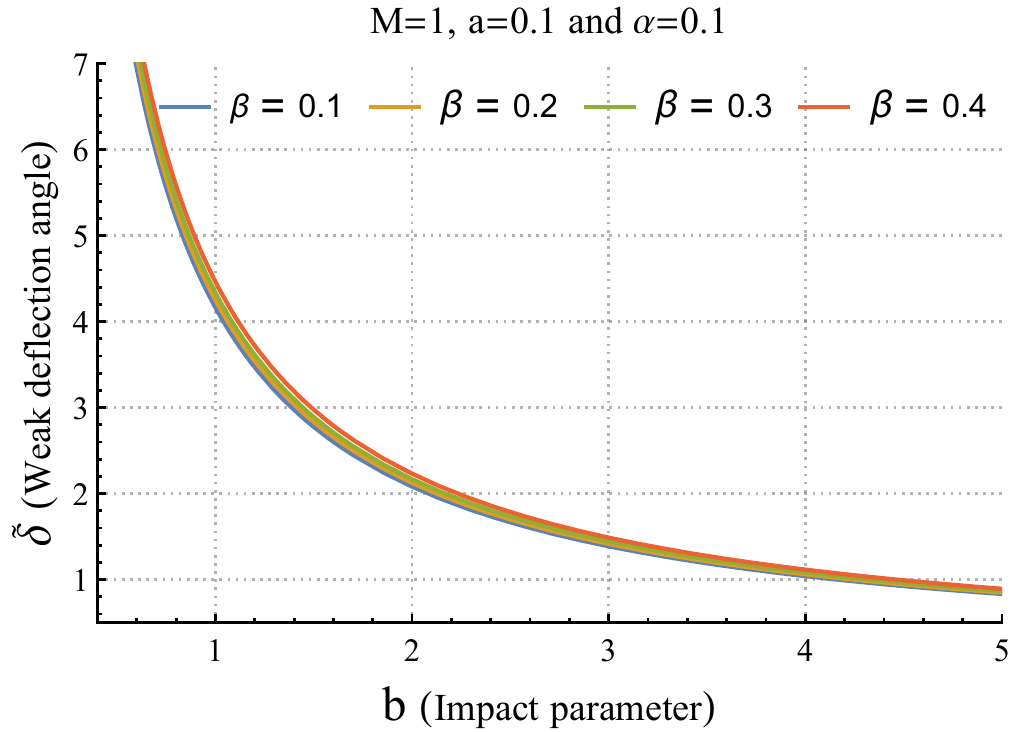}}}
    \caption{The profile of the deflection angle around the GUP corrected black with a background of the cloud of string field as a function of impact parameter $b$ for different values of $a$ (top left panel), $\alpha$ (top right panel) and $\alpha$ (bottom panel). }
    \label{fig:def_an}
\end{figure*}
It is clearly seen from the expression that the deflection angle can be affected implicitly by the effects of string cloud field parameter $a$ and the dimensionless quantum correction parameters $\alpha$ and $\beta$. With this in view, working within the GBT method one is able to determine the deflection angle for any asymptotically flat optical metric as that of its particular properties. It is now important to gain a deeper understanding in relation to how these black hole parameters can influence on the deflection angle in the weak approximation. For that, in Fig~\ref{fig:def_an} we show the profile of the deflection angle as a function of the impact parameter $b$. The top left and right panels respectively refer to the impact of the string cloud $a$ and the dimensionless quantum correction parameter $\alpha$, while the bottom panel to the impact of the dimensionless parameter $\beta$ on the deflection angle profile for possible various cases, as seen in Fig~\ref{fig:def_an}. We can observe from Fig.~\ref{fig:def_an} that the deflection angle decreases as $b$ increases. It is to be emphasized that the changing rate of the deflection is similar and always increases as a consequence of the effects of $a$ and $\beta$ parameters. It must then be noted that  $a$ and $\beta$ parameters can be interpreted as repulsive gravitational charges of GUP corrected black hole with the string cloud field, thus resulting in shifting the deflection angle upward to its larger values. Interestingly, it is found that the dimensionless parameter $\alpha$ has the opposite physical impact in contrast to $a$ and $\beta$, which can slightly shift the deflection angle down toward to its lower values. However, the impact of parameter $\alpha$ is comparable to the impact of the parameter $\beta$ on the deflection angle. Unlike $a$ and $\beta$ it is obvious that $\alpha$ has the impact that makes the deflection angle decrease, thereby being consistent with the physical interpretation of a black hole parameter as an attractive gravitational charge. To be more informative we further consider other various physical phenomena to gain a deeper understanding of the remarkable properties of the spacetime of GUP-corrected black holes. This is what we intend to examine in the next
sections. 
 
\section{Quasinormal modes}
In this section, we will consider the GUP-corrected black holes with the cloud of string to investigate QNMs. In general, non-rotating black hole perturbations are governed by a Schrödinger-like wave equation of the form:
\begin{equation}
\frac{\text{d}^{2}\Psi}{\text{d}r_{\ast }^{2}}+\left( \omega^{2}-V(r)\right) \Psi=0,  \label{s3}
\end{equation}
where $V(r)$ is the effective potential of the QNMs. Assuming the perturbations depend on
time as $e^{-i\omega t}$. Then,  the imaginary part of $\omega$ must be
negative in order to have damping. The tortoise coordinate $r_{\ast }$ is defined by $\frac{d}{dr_{\ast }}
=A\frac{d}{dr}$. The effective potentials are
\begin{equation}
V_{0}=\frac{\left( \frac{1}{2}+l\right) }{r^{2}}A+\frac{A^{\prime }}{r} \label{pots}
\end{equation}%
\begin{equation}
V_{1}=\frac{\left( \frac{1}{2}+l\right) }{r^{2}}A   \label{pots10} 
\end{equation}
\begin{equation}
V_{2}=\frac{\left( \frac{1}{2}+l\right) }{r^{2}}\left( \left( \frac{1}{%
2}+l\right) A\pm \frac{r\sqrt{A}A^{\prime }}{2}\mp A^{3/2}\right) .
\label{pot11}
\end{equation}
where $V_0$ describes a massless scalar field, $V_1$
for electromagnetic (EM) field, $V_2$ describes massless Dirac
field and $l$ is the standard spherical harmonics indices. 
To explore the impact of the model black hole parameters, graphs of the massless scalar field potential (\ref{pots}) are made. The following Figure \ref{pot1} will only look at scalar perturbations because all three potentials (\ref{pots}), (\ref{pots10}) and (\ref{pot11}) behave identically in terms of black hole parameters.  
\begin{figure}[H]
\begin{center}
\includegraphics[scale=0.75]{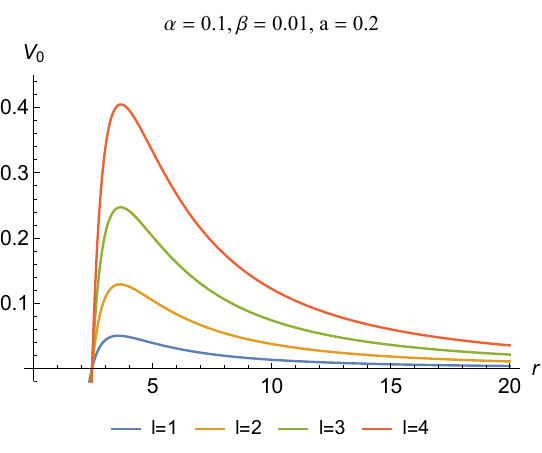}
\includegraphics[scale=0.75]{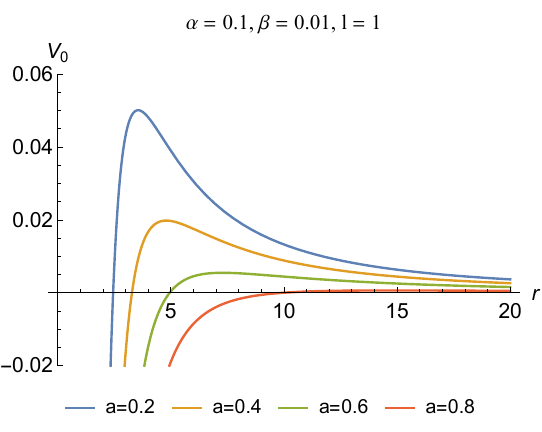}
\includegraphics[scale=0.75]{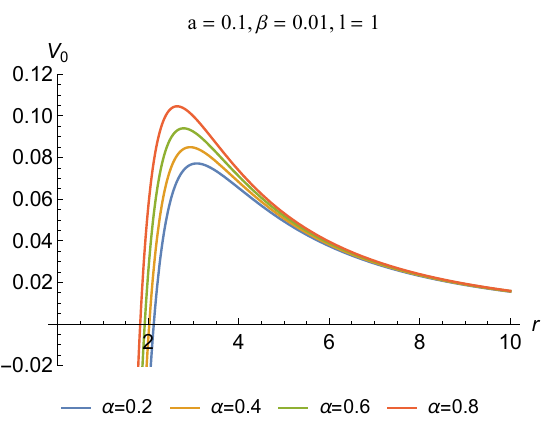}
\includegraphics[scale=0.75]{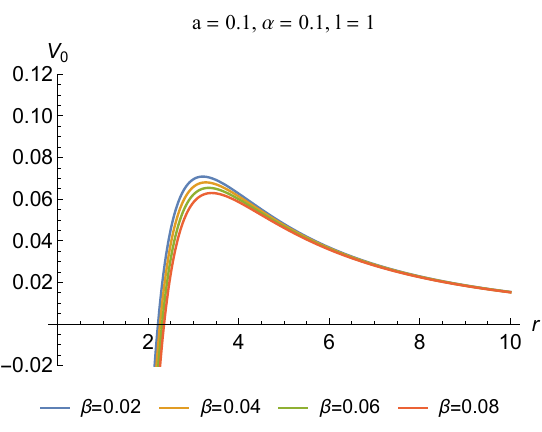}
\end{center}
\caption{Variation of the scalar potential  with the radial distance $r$ is plotted for various values of  black hole model parameters}\label{pot1}
\end{figure}
In the top of Fig. \ref{pot1}, we plotted the massless scalar potential for various multipole moment $l$ values and the string parameter $a$. As expected, the peak of the potential rises as $ l$ grows while increasing the value of $a$ decreases the potential curve considerably. In the bottom of Fig. \ref{pot1}, we illustrate how the massless scalar potential varies with different values of the GUP parameters $\alpha$ and $\beta$. It is clear that as the value of $\alpha$ increases, so does the potential curve. However, increasing $\beta$ produces the reverse effect.
Therefore, the cloud string parameter and GUP parameter $\alpha$ are expected to behave differently in the QNMs spectrum. \\ Our next step is to investigate the QNMs of the corrected GUP Schwarzschild string cloud black hole metric (\ref{equ9}) using the 6th-order WKB approximation \cite{Iyer, Konoplya} approach with a larger value of $l$. The reason for selecting a relatively greater value of multipole moment $l$ is that the error associated with the WKB approach is greatly reduced for higher $l$. The results are presented in Tables \ref{taba1}, \ref{taba2} and \ref{taba3}.
\begin{center}
\begin{tabular}{|c|c|c|c|}
 \hline 
         
   \multicolumn{4}{|c|}{ $l=2$, $n=0$, $M=1$, $\alpha=0.1$ and $\beta =0.01$}
\\ \hline  
         
   $a$ & Scalar field  &  EM field &  Fermionic field
\\ \hline 
$0$ & $0.49343-0.091219i$ & $0.46641-0.08811i$ & $0.48766-0.096593i$ \\ 
$0.1$ & $0.41956-0.074219i$ & $0.39888-0.07204i$ & $0.41590-0.078022i$ \\ 
$0.2$ & $0.35013-0.058917i$ & $0.33479-0.05745i$ & $0.34814-0.061494i$ \\ 
$0.3$ & $0.28535-0.045327i$ & $0.27440-0.04439i$ & $0.28461-0.046980i$ \\ 
$0.4$ & $0.22545-0.033470i$ & $0.21802-0.03290i$ & $0.22557-0.034457i$ \\ 
$0.5$ & $0.17073-0.023364i$ & $0.16604-0.02305i$ & $0.17137-0.023899i$ \\ 
$0.6$ & $0.12160-0.015034i$ & $0.11893-0.01488i$ & $0.12246-0.015285i$ \\ 
$0.7$ & $0.07861-0.008504i$ & $0.07731-0.00844i$ & $0.07942-0.008598i$ \\ 
$0.8$ & $0.04258-0.003801i$ & $0.04211-0.00378i$ & $0.04317-0.003825i$%
\\ 
 \hline
\end{tabular}
\captionof{table}{The QNM of the GUP Schwarzschild 
 string cloud black hole  for scalar, EM, and fermionic perturbations for various string cloud parameter $a$.} \label{taba1}
\end{center}
\begin{center}
\begin{tabular}{|c|c|c|c|} \hline 
         
   \multicolumn{4}{|c|}{ $l=2$, $n=0$, $M=1$, $a=0.1$ and $\beta =0.01$}
\\ \hline 
         
   $\alpha$ & Scalar field  &  EM field &  Fermionic field
\\ \hline 
$0$ & $0.41013-0.07255i$ & $0.38991-0.07042i$ & $0.40655-0.07626i$ \\ 
$0.1$ & $0.41956-0.07421$ & $0.39888-0.07204i$ & $0.41590-0.07802i$ \\ 
$0.2$ & $0.42944-0.07596i$ & $0.40827-0.073739i$ & $0.42569-0.07985i$ \\ 
$0.3$ & $0.43979-0.07779i$ & $0.41811-0.075516i$ & $0.43595-0.08178i$ \\ 
$0.4$ & $0.45065-0.07971i$ & $0.42843-0.077381i$ & $0.44672-0.08380i$ \\ 
$0.5$ & $0.46206-0.08173i$ & $0.43928-0.079341i$ & $0.45803-0.08592i$ \\ 
$0.6$ & $0.47407-0.08386i$ & $0.45070-0.08140i$ & $0.46993-0.08815i$ \\ 
$0.7$ & $0.48671-0.08609i$ & $0.46272-0.08357i$ & $0.48247-0.09051i$ \\ 
$0.8$ & $0.50005-0.08845i$ & $0.47540-0.08586i$ & $0.49569-0.09299i$%
\\ 
 \hline
\end{tabular}
\captionof{table}{The QNM of the GUP Schwarzschild 
 string cloud black hole for scalar, EM, and fermionic perturbations for various GUP parameters $\alpha$.} \label{taba2}
\end{center}

\begin{figure}
    \centering
    \includegraphics[width=18cm]{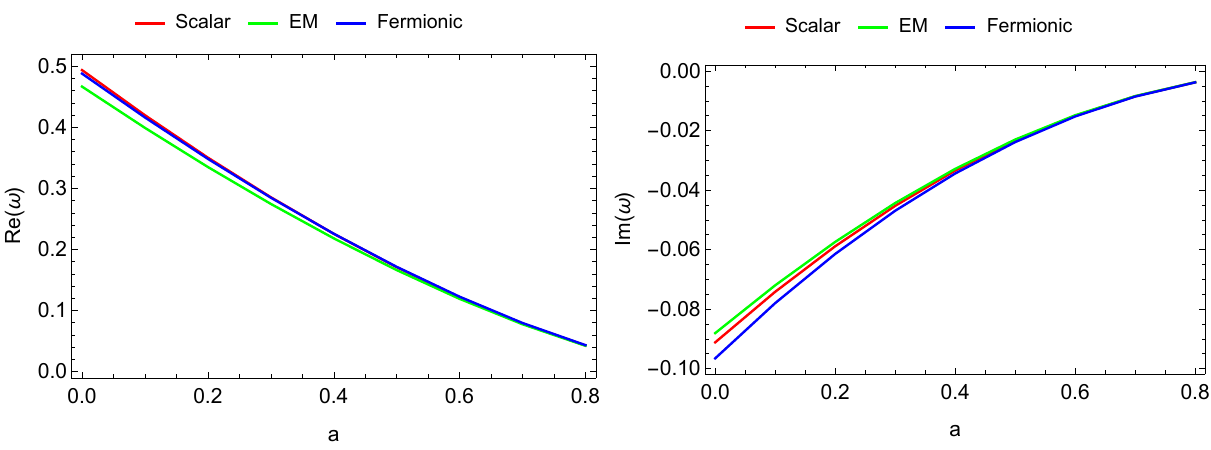}
    \caption{Real (left) and imaginary (right ) parts of the QNMs of the GUP corrected black hole for scalar field, EM field, and fermionic field perturbations as a function of the cloud string parameter $a$.}
    \label{realA}
\end{figure}

\begin{figure}
    \centering
    \includegraphics[width=18cm]{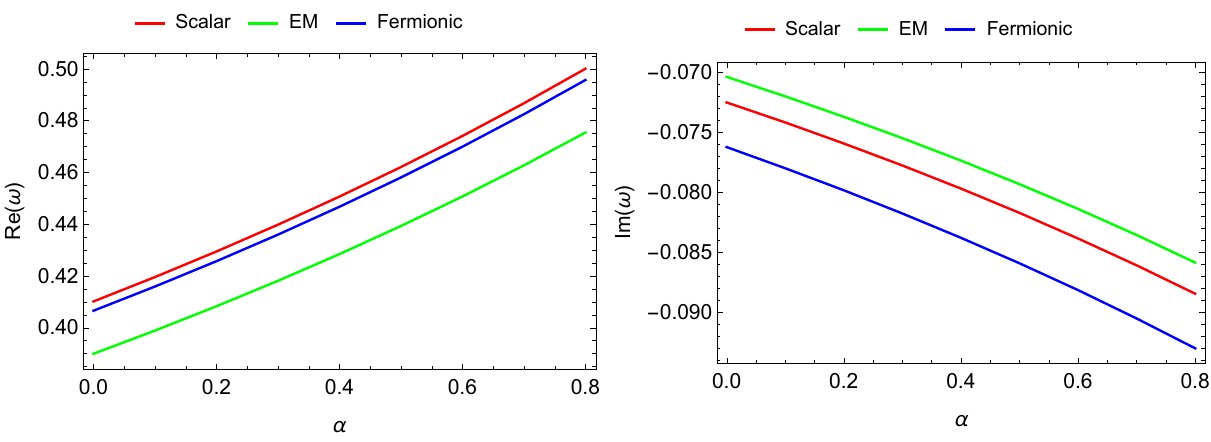}
    \caption{Real (left) and imaginary (right ) parts of the QNMs of the GUP corrected black hole for scalar field, EM field, and fermionic field perturbations as a function of the GUP parameter $\alpha$.}
    \label{realAph}
\end{figure}

\begin{figure}
    \centering
    \includegraphics[width=18cm]{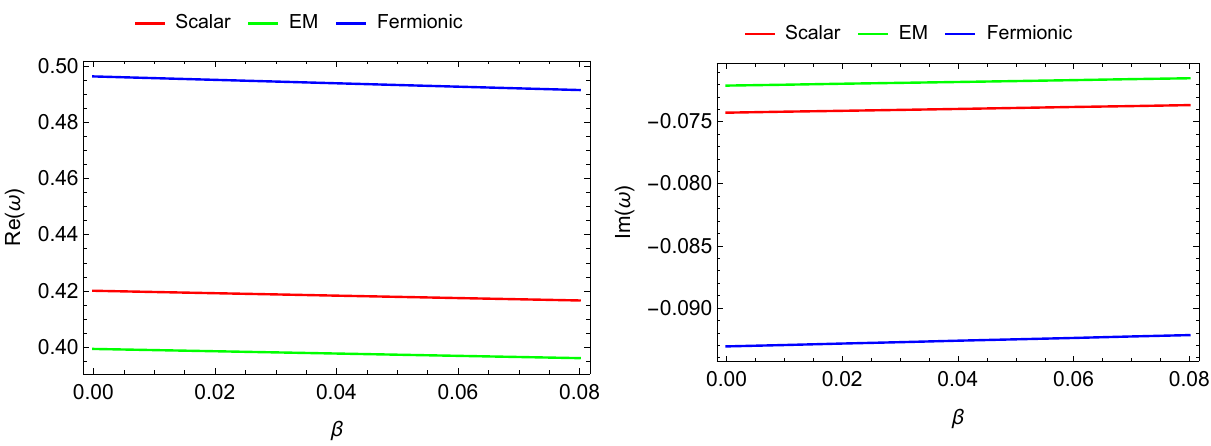}
    \caption{Real (left) and imaginary (right ) parts of the QNMs of the GUP corrected black hole for scalar field, EM field, and fermionic field perturbations as a function of the GUP parameter $\beta$.}
    \label{realB}
\end{figure}

\begin{figure}
    \centering
    \includegraphics[scale=0.75]{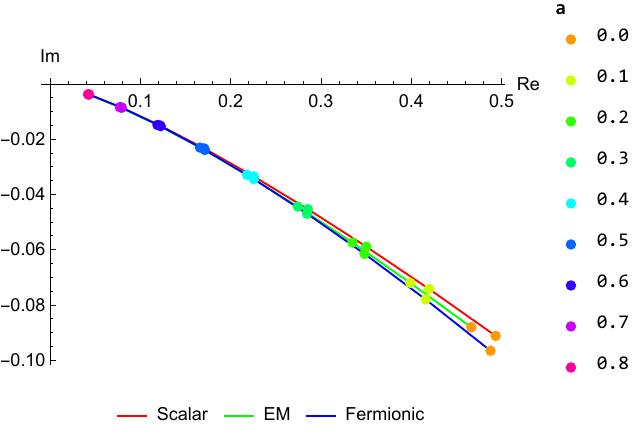}
\includegraphics[scale=0.75]{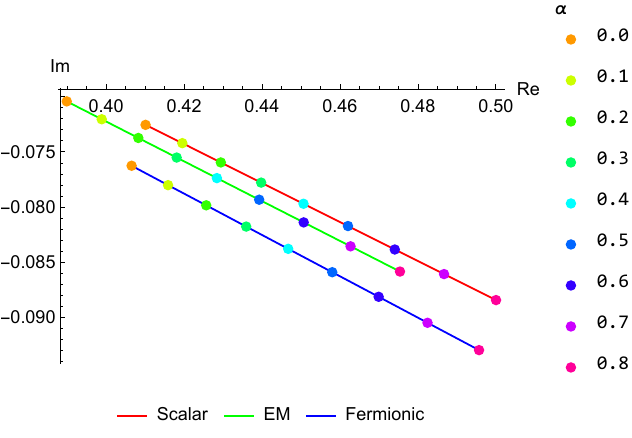}
    \caption{Complex frequency plane for the three perturbations  showing the behavior of the quasinormal frequencies}
    \label{imagA}
\end{figure}

\begin{center}
\begin{tabular}{|c|c|c|c|} \hline 
         
   \multicolumn{4}{|c|}{ $l=2$, $n=0$, $M=1$, $\alpha=0.1$ and $a =0.1$}
\\\hline 
         
   $\beta$ & Scalar field  &  EM field &  Fermionic field
\\ \hline 
$0$ & $0.42000-0.07429i$ & $0.39929-0.07211i$ & $0.49630-0.09310i$ \\ 
$0.01$ & $0.41956-0.07421i$ & $0.39888-0.07204i$ & $0.49569-0.09299i$ \\ 
$0.02$ & $0.41913-0.07414i$ & $0.39847-0.07196i$ & $0.49508-0.09287i$ \\ 
$0.03$ & $0.4187-0.074066i$ & $0.39806-0.07189i$ & $0.49447-0.09276i$ \\ 
$0.04$ & $0.41826-0.07399i$ & $0.39765-0.07182i$ & $0.49386-0.09264i$ \\ 
$0.05$ & $0.41783-0.07391i$ & $0.39724-0.07174i$ & $0.49325-0.09253i$ \\ 
$0.06$ & $0.41740-0.07383i$ & $0.39683-0.07167i$ & $0.49265-0.09242i$ \\ 
$0.07$ & $0.41697-0.07376i$ & $0.39642-0.07159i$ & $0.49205-0.09230i$ \\ 
$0.08$ & $0.41654-0.07368i$ & $0.39601-0.07152i$ & $0.49145-0.09219i$%
\\ 
 \hline
\end{tabular}
\captionof{table}{The QNM of the GUP Schwarzschild 
 string cloud black hole for scalar, EM, and fermionic perturbations for various GUP parameters $\beta$.} \label{taba3}
\end{center}
 To see how the model parameters affect the QNM spectrum, we directly plotted the real and imaginary QNMs against the model parameters. For this objective, Fig. \ref{realA} shows the variation of real and imaginary QNMs with respect to the cloud string parameter $a$ for the three types of perturbations. Increasing $a$ leads to a considerable decrease in real QNMs or oscillation frequencies for all three perturbations. The fluctuation is nearly linear. Similarly, when $a$ increases, the damping rate or decay rate of gravitational waves raises considerably so, the perturbations decay slowly. The damping rate varies roughly linearly with respect to $a$.\\ In Fig. \ref{realAph} we show the variation of real and imaginary
QNMs with respect to the GUP parameter $\alpha$. Increasing the GUP parameter $\alpha$ significantly increases real QNMs or oscillation frequencies across all three perturbations. The fluctuations are roughly linear. Increasing $\alpha$ leads to a significant drop in the damping or decay rate of gravitational waves, so the perturbations decay faster with increasing  $\alpha$. According to our research, both parameters i.e. string cloud $a$ and GUP $\alpha$ have distinct and different impacts on QNMs. \\Figure \ref{realB}  shows the variation of real and imaginary
QNMs with respect to the GUP parameter $\beta$. It is seen that the effect is very small on both oscillation and damping rates. It was found that increasing the value of parameters $a$ and $\beta$ reduce the relative change of the effective potential, resulting in lower peak values and a drop in gravitational wave oscillation frequency and damping. The converse effect is observed when $\alpha$ is increased. Thus, the findings confirm the effect of model parameters on QNMs that we saw in the potential graphs (Fig. \ref{pot1}).\\
In Fig. \ref{imagA}, We plot the QNM frequencies in the complex frequency plane. Figures are consistent with previous Figs. \ref{realA} and \ref{realAph}.

\section{Evolution of perturbations}
\begin{figure}[H]
\begin{center}
\begin{tabular}{lcr}
\includegraphics[scale=0.45]{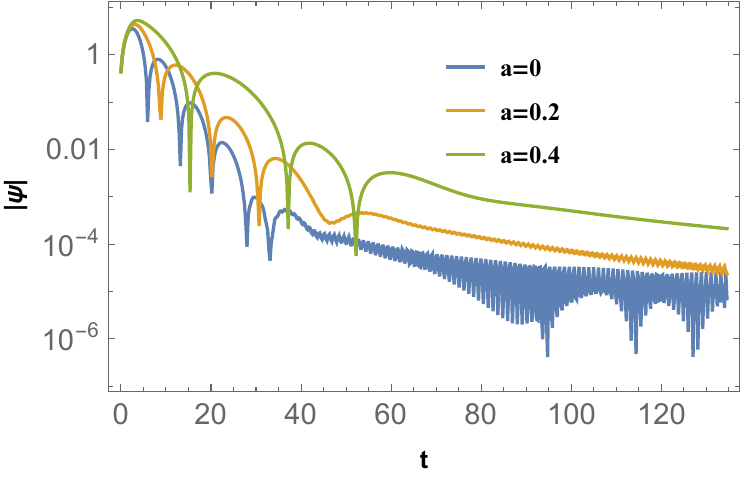}&
\includegraphics[scale=0.45]{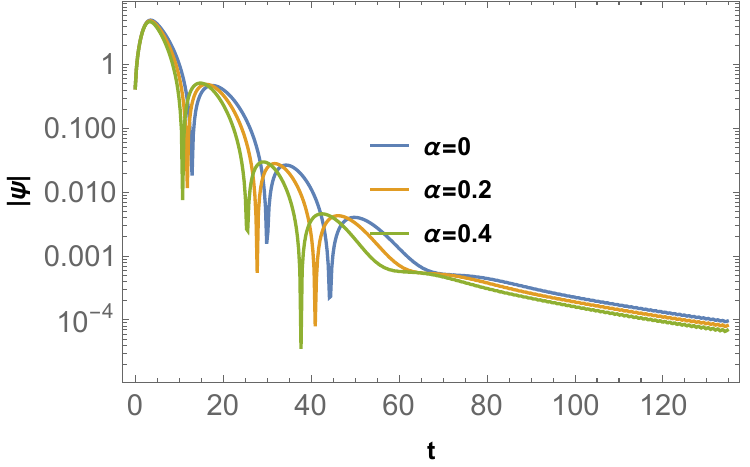}&
\includegraphics[scale=0.45]{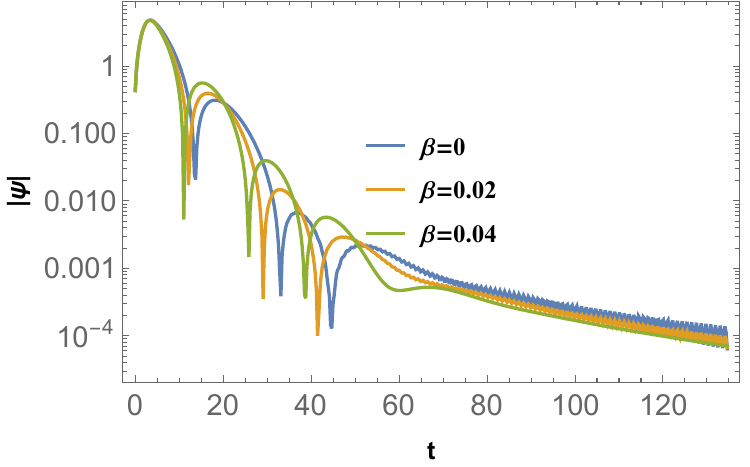}
\end{tabular}
\end{center}
\caption{Ringdown waveforms for electromagnetic perturbation. The left one is for different values of $a$ with $\alpha=0.3$, $\beta=0.03$. The middle one is for different values of $\alpha$ with $a=0.3$, $\beta=0.03$. The right one for different values of $\beta$ with $a=\alpha=0.3$. We have taken $\ell=2$ for all three cases.}\label{ringem}
\end{figure}

\begin{figure}[H]
\begin{center}
\begin{tabular}{lcr}
\includegraphics[scale=0.45]{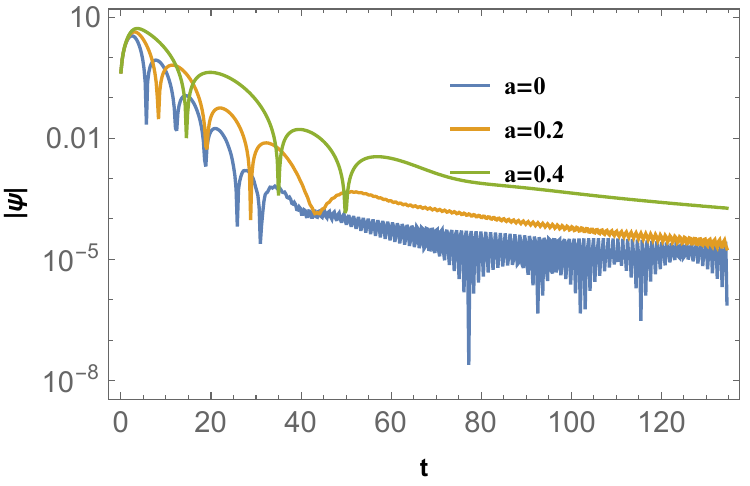}&
\includegraphics[scale=0.45]{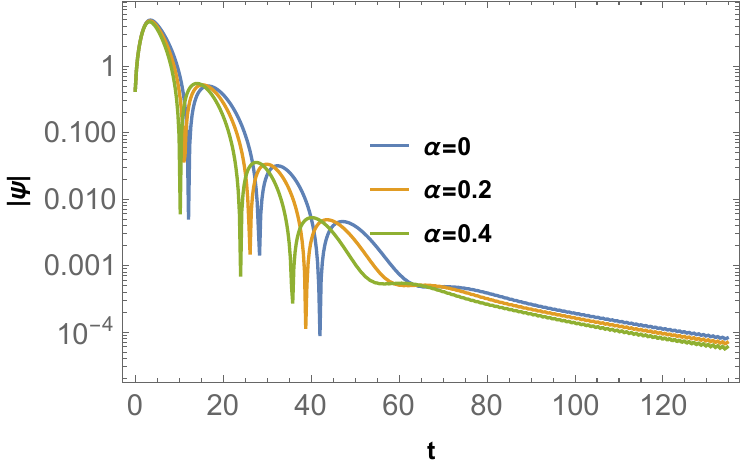}&
\includegraphics[scale=0.45]{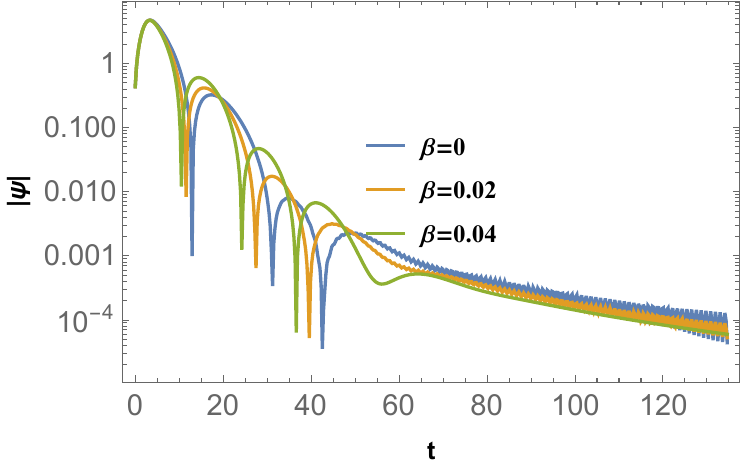}
\end{tabular}
\end{center}
\caption{Ringdown waveforms for scalar perturbation. The left one is for different values of $a$ with $\alpha=0.3$, $\beta=0.03$. The middle one is for different values of $\alpha$ with $a=0.3$, $\beta=0.03$. The right one for different values of $\beta$ with $a=\alpha=0.3$. We have taken $\ell=2$ for all three cases.}\label{ringscalar}
\end{figure}

\begin{figure}[H]
\begin{center}
\begin{tabular}{lcr}
\includegraphics[scale=0.45]{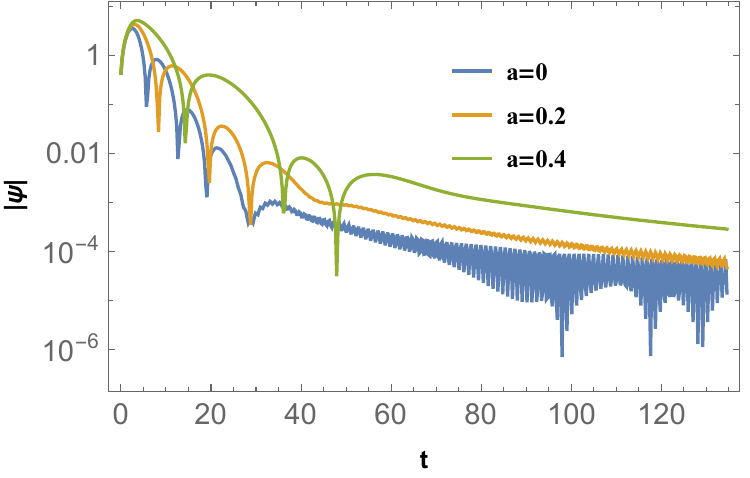}&
\includegraphics[scale=0.45]{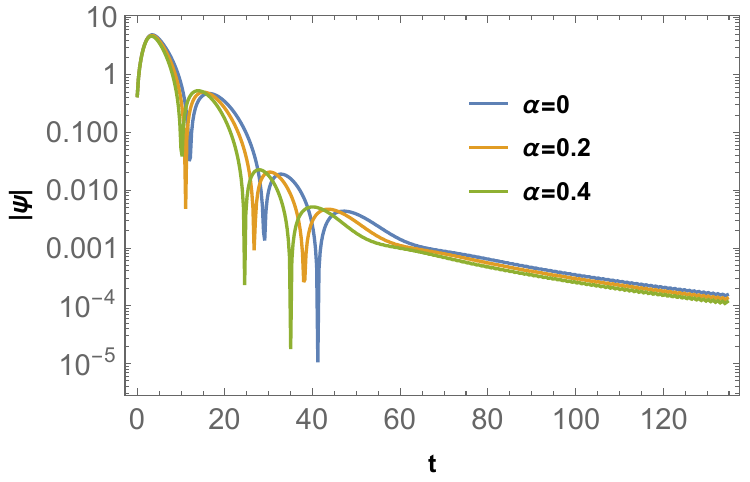}&
\includegraphics[scale=0.45]{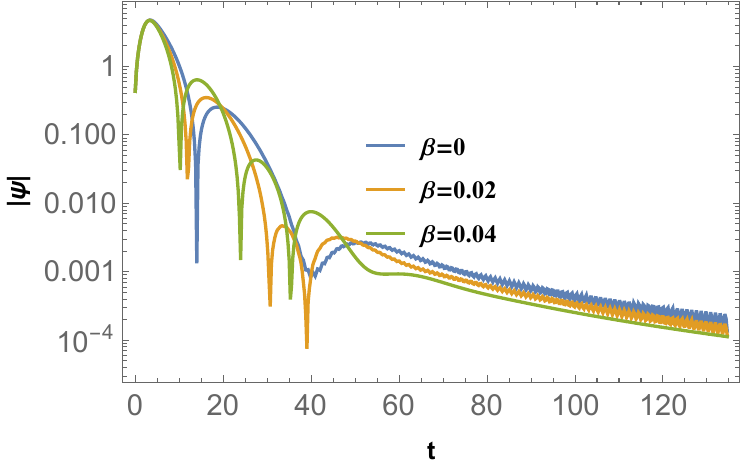}
\end{tabular}
\end{center}
\caption{Ringdown waveforms for fermionic perturbation. The left one is for different values of $a$ with $\alpha=0.3$, $\beta=0.03$. The middle one is for different values of $\alpha$ with $a=0.3$, $\beta=0.03$. The right one for different values of $\beta$ with $a=\alpha=0.3$. We have taken $\ell=2$ for all three cases.}\label{ringfermion}
\end{figure}

It would be instructive to study ringdown waveforms to gauge the effect of model parameters. Since they bear the signature of the underlying spacetime, it is imperative to study them. To this end, we need to solve the time-dependent wave equation. We employed the time domain integration method \cite{gundlach1}. We used initial conditions $\Psi(r_*,t) = \exp \left[ -\dfrac{(r_*-\hat{r}_{*})^2}{2\sigma^2} \right]$ and $\Psi(r_*,t)\vert_{t<0} = 0$ with $r_*=5$, $\hat{r}_*=0.4$. $\Delta t$ and $\Delta r_{*}$ are taken so that their values are in congruence with the Von Neumann stability condition.\\Time domain profiles of electromagnetic, scalar, and fermionic perturbations are shown in Fig. [\ref{ringem}, \ref{ringscalar}, \ref{ringfermion}], respectively. These figures further verify the effect of model parameters on the QNMs that we observed in the previous section.

\section{Sparsity of Hawking emission}
The impact of the string parameter a and GUP parameters $\alpha$ and $\beta$ can further be probed with the help of the greybody factor and sparsity of the Hawking radiation. The greybody factor is the transmission probability of Hawking radiation, whereas the sparsity provides information regarding the continuous flow of Hawking radiation.

\subsection{ Greybody factor}

In this part of our inquiry, we apply the method described in \cite{gfs1,gfs2,gfs71, gfs3, gfs72} to investigate rigorous bounds on greybody factors. Because this part of our analysis demonstrates that all three perturbations behave similarly in terms of greybody factors, we will only consider scalar perturbations in the following part.  According to \cite{gfs1,gfs3}, the following formula can be used to obtain the rigorous bound of the greybody factor:\begin{equation}
T\left( w\right) \geq \sec h^{2}\left( \frac{1}{2w}%
\int_{r_{h}}^{+\infty }V\text{d}r_{\ast }\right) .  \label{is8}
\end{equation}
The greybody factor of a massless scalar field will be calculated using the effective potential given in Eq. (\ref{pots}).  
 Therefore, Eq. (\ref{is8}) becomes
\begin{equation}
T\left( w\right) \geq \sec h^{2}\left( \frac{1}{2\omega }
\int_{r_{h}}^{\infty }\left( \frac{l\left( l+1\right) }{r^{2}}+\frac{A^{\prime }}{r}\right) \text{d}r \right) .\label{in10}
\end{equation}
The analytical solution is given by  
\begin{equation}
T\left( w\right) \geq \sec h^{2}\frac{1}{\omega }\left[ \left( -\frac{%
8M^{2}+2M(4lr_h +4l^2 r_h
-(1-a) \alpha) +(1-a)^2 \beta}{8Mr_{h}^{2}}\right) \right]. 
 \label{gf1}
\end{equation}
\begin{figure}[H]
\begin{center}
\begin{tabular}{lcr}
\includegraphics[scale=0.55]{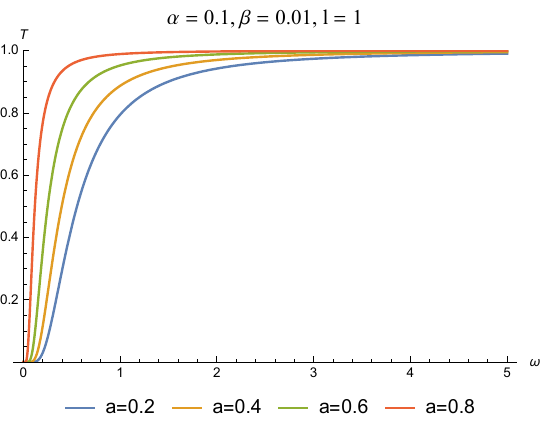}&
\includegraphics[scale=0.55]{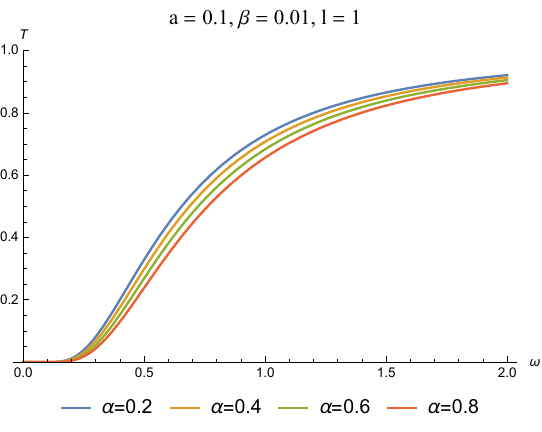}&
\includegraphics[scale=0.55]{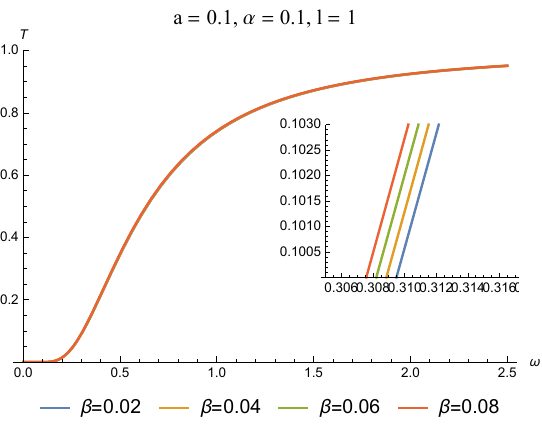}
\end{tabular}
\end{center}
\caption{The greybody bounding of massless scalar field for various values of  black hole parameters}\label{gf123}
\end{figure}
The greybody bounding of massless scalar field is plotted for various values of black hole parameters in Fig. \ref{gf123}. The accompanying graphs show that as the string cloud parameter $a$ increases, so does the greybody factor's bound. This observation shows that the present string cloud exhibits poor barrier characteristics and has larger greybody factors. The GUP parameter $\beta$ follows a similar trend, whereas the GUP parameter $\alpha$ follows an opposite trend. This is because increasing the value of the GUP parameter $\alpha$ increases the relative change in the effective potential, allowing less thermal radiation to reach the observer at spatial infinity.

\subsection{ Sparsity of the Hawking radiation}
\begin{figure}[H]
\begin{center}
\begin{tabular}{lcr}
\includegraphics[scale=0.45]{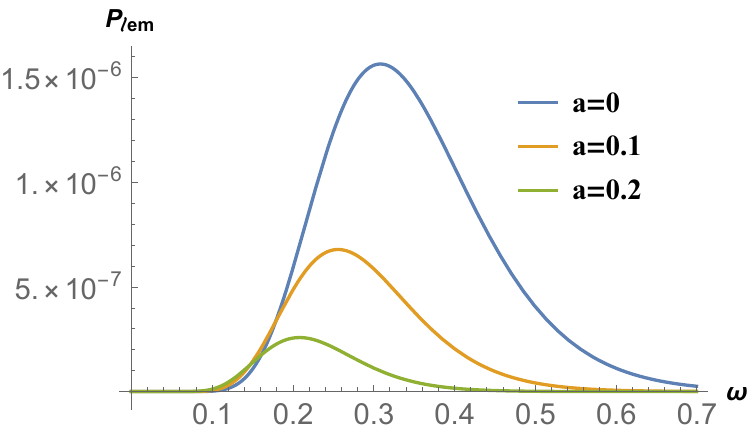}&
\includegraphics[scale=0.45]{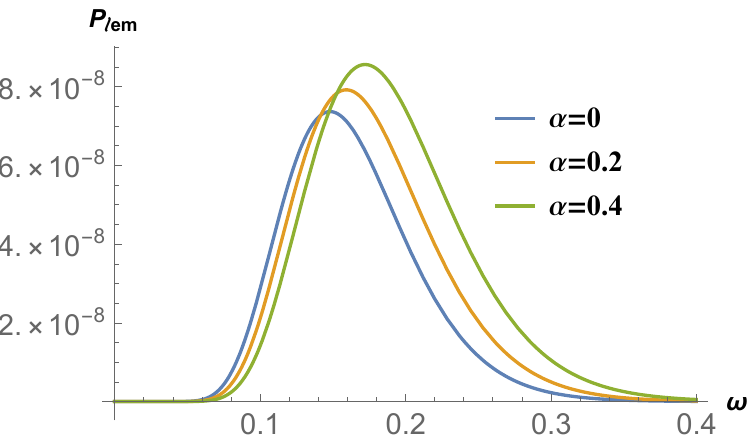}&
\includegraphics[scale=0.45]{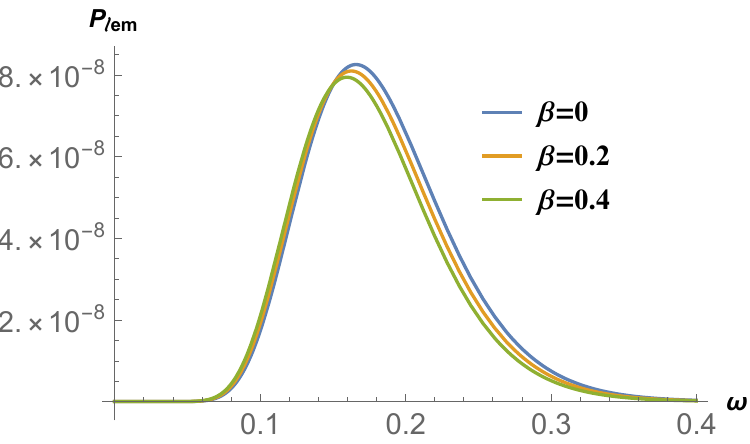}
\end{tabular}
\end{center}
\caption{Power spectrum of the black hole for electromagnetic perturbation. The left one is for different values of $a$ with $\alpha=0.3$, $\beta=0.03$. The middle one is for different values of $\alpha$ with $\alpha=0.3$, $\beta=0.03$. The right one for different values of $\beta$ with $a=\alpha=0.3$. We have taken $\ell=1$ for all three cases.}\label{emfig}
\end{figure}

\begin{figure}[H]
\begin{center}
\begin{tabular}{lcr}
\includegraphics[scale=0.45]{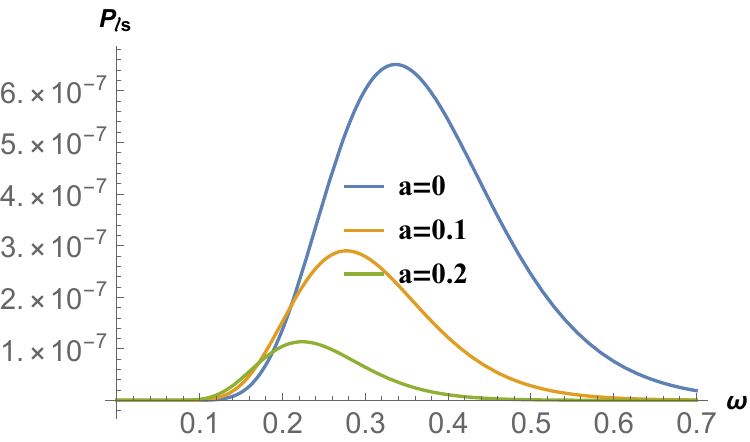}&
\includegraphics[scale=0.45]{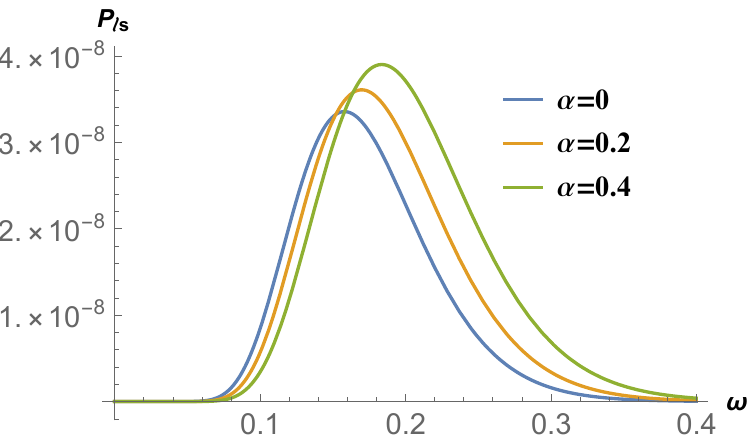}&
\includegraphics[scale=0.45]{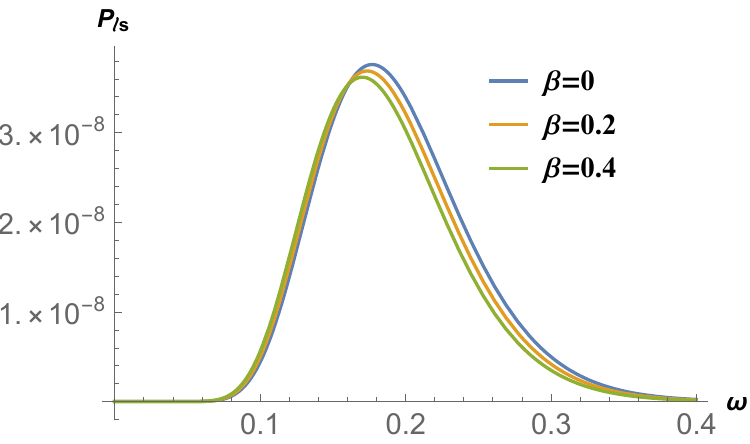}
\end{tabular}
\end{center}
\caption{Power spectrum of the black hole for scalar perturbation. The left one is for different values of $a$ with $\alpha=0.3$, $\beta=0.03$. The middle one is for different values of $\alpha$ with $\alpha=0.3$, $\beta=0.03$. The right one for different values of $\beta$ with $a=\alpha=0.3$. We have taken $\ell=1$ for all three cases.}
\label{scalarfig}
\end{figure}

\begin{figure}[H]
\begin{center}
\begin{tabular}{lcr}
\includegraphics[scale=0.45]{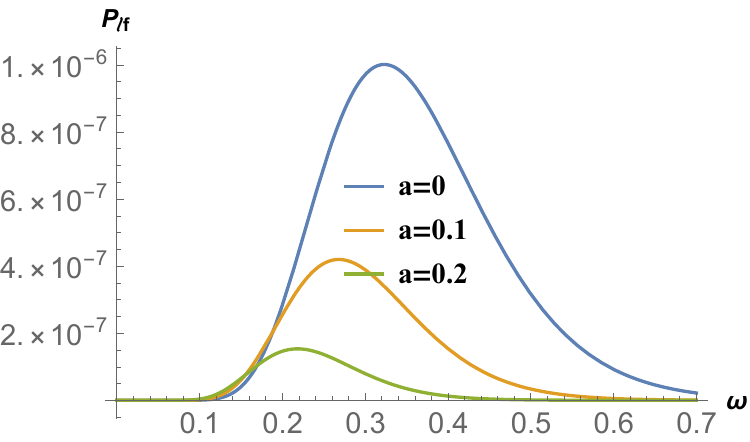}&
\includegraphics[scale=0.45]{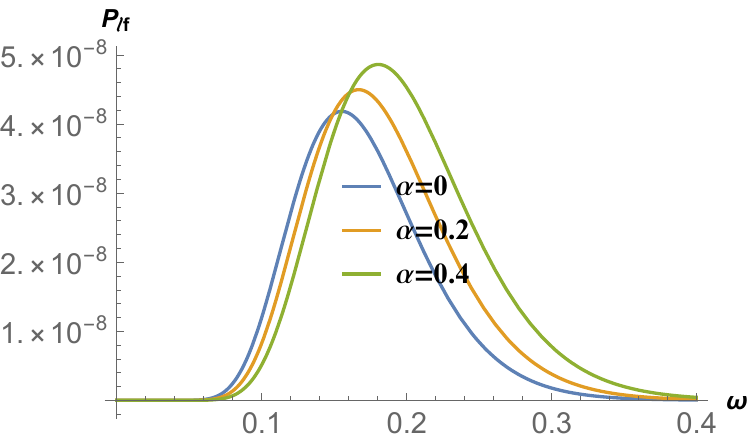}&
\includegraphics[scale=0.45]{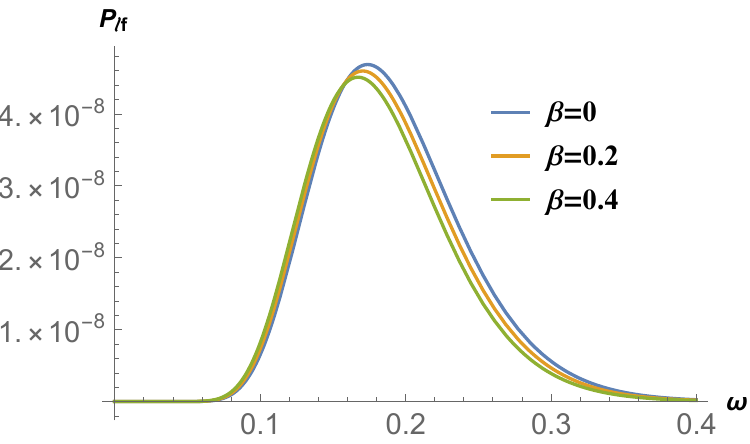}
\end{tabular}
\end{center}
\caption{Power spectrum of the black hole for fermionic perturbation. The left one is for different values of $a$ with $\alpha=0.3$, $\beta=0.03$. The middle one is for different values of $\alpha$ with $\alpha=0.3$, $\beta=0.03$. The right one for different values of $\beta$ with $a=\alpha=0.3$. We have taken $\ell=1$ for all three cases.}\label{fermionicfig}
\end{figure}

We, in this section, study the Hawking spectrum and sparsity for the BH under consideration. The total power emitted by a BH as massless particles is given by  \cite{yg2017, fg2016}
\begin{equation}\label{ptot}
P_{tot}=\sum_\ell \int_{0}^{\infty} P_\ell\left(\omega\right) \text{d}\omega.
\end{equation}
The power spectrum in the $\ell th$ mode is 
\begin{equation}\label{pl}
P_\ell\left(\omega\right)=\frac{A}{8\pi^2}T(\omega)\frac{\omega^3}{e^{\omega/T_{H}}-1},
\end{equation}
where $T(\omega)$ is the greybody factor and $A$ is taken to be the horizon area \cite{yg2017}. The Hawking temperature is 
\begin{equation}
    T_H=\frac{1}{4\pi}\frac{dA(r)}{dr}=\frac{(a-1)^2 M}{\pi  (-a \beta +4 M ((a-1) \alpha +2 M)+\beta )}.
\end{equation}
Now, we can explore the impact of string and GUP parameters on the power spectrum of Hawking radiation. Figs. [\ref{emfig}, \ref{scalarfig}, \ref{fermionicfig}] illustrate the effect of $a, \alpha$, and $\beta$ on $P_{\ell}$. 
We can infer from Fig. \ref{emfig} that the parameters a and $\beta$ adversely impact the power emitted. In contrast, the parameter $\alpha$ has a favorable impact on the power spectrum of electromagnetic perturbation. However, the variation of power emitted with $\beta$ is marginal. The position of the peak, $\omega_{max}$, shifts towards the left as we increase either a or $\beta$ but shifts towards the right as we increase $\alpha$. The same conclusions are true for scalar and fermionic perturbations, as evident in Figs. \ref{scalarfig} and  \ref{fermionicfig}. We next define the sparsity of Hawking radiation as 
\begin{equation}
\eta=\frac{\tau_{gap}}{\tau_{emission}},
\label{eta}
\end{equation}
where $\tau_{gap}=\frac{\omega_{max}}{P_{tot}}$ is the time gap between successive radiation quanta and $\tau_{emission}$ is taken to be the time period for the radiation with frequency $\omega_{max}$. The parameter $\eta$ signifies continuous or discontinuous flow of radiation. A much larger value of $\eta$ implies a discontinuous flow of radiation. Numerical values of sparsity $\eta$ and related quantities are shown in Tables \ref{a}, \ref{alpha} and  \ref{beta}. 
\begin{table}[!htp]
\centering
\caption{Numerical values of $\omega_{max}$, $P_{max}$, $P_{tot}$, and $\eta$ for various values of $a$ with $\alpha=0.3$, $\beta=0.03$, and $\ell=1$.}
\setlength{\tabcolsep}{-.2mm}
\begin{tabular}{|c|c|c|c|c|c|c|}
\hline
Perturbation & a & 0 & 0.15 & 0.30 & 0.45 & 0.60  \\
\hline
\multirow{4}{*}{Scalar}& $ \omega _{max }$ & 0.33673 & 0.246756 & 0.172809 & 0.110621 & 0.0721875 \\
& $ P_{max }$ & 6.50689$\times 10^{-7}$ & 1.84367$\times 10^{-7}$ & 3.73532$\times 10^{-8}$ & 4.37182$\times 10^{-9}$ & 1.88560$\times 10^{-10}$ \\
& $ P_{\text{tot}}$ & 1.62626$\times 10^{-7}$ & 3.32368$\times 10^{-8}$ & 4.59117$\times 10^{-9}$ & 3.39813$\times 10^{-10}$ & 8.18471$\times 10^{-12}$ \\
& $ \eta$ & 110967. & 291565. & 1.03521$\times 10^6$ & 5.73138$\times 10^6$ & 1.01331$\times 10^8$ \\
\hline
\multirow{4}{*}{Electromagnetic}& $ \omega _{max }$ & 0.308339 & 0.232128 & 0.167901 & 0.110621 & 0.0721875 \\
& $ P_{max }$ &$ 1.56292\times 10^{-6}$ & 4.25853$\times 10^{-7}$ & 8.21581$\times 10^{-8}$ & 9.15159$\times 10^{-9}$ & 3.38828$\times 10^{-10}$ \\
& $ P_{\text{tot}}$ & 3.76261$\times 10^{-7}$ & 7.42056$\times 10^{-8}$ & 9.80295$\times 10^{-9}$ & 6.85346$\times 10^{-10}$ & 1.53128$\times 10^{-11}$ \\
& $ \eta$ & 40215. & 115569. & 457686. & 2.84177$\times 10^6$ & 5.41613$\times 10^7$ \\
\hline
\multirow{4}{*}{Fermionic}& $ \omega _{max }$ & 0.321204 & 0.243245 & 0.172809 & 0.110621 & 0.0721875 \\
& $ P_{max }$ & 1.00065$\times 10^{-6} $& 2.58743$\times 10^{-7}$ & 4.66908$\times 10^{-8}$ & 4.67612$\times 10^{-9}$ & 1.62830$\times 10^{-10}$ \\
& $ P_{\text{tot}}$ & 2.45724$\times 10^{-7}$ & 4.5999$\times 10^{-8}$ & 5.67877$\times 10^{-9}$ & 3.61711$\times 10^{-10}$ & 7.02301$\times 10^{-12}$ \\
& $ \eta$ & 66824.1 & 204720. & 836949. & 5.3844$\times 10^6$ & 1.18092$\times 10^8$ \\
\hline
\end{tabular}
\label{a}
\end{table}
\begin{table}[!htp]
\centering
\caption{Numerical values of $\omega_{max}$, $P_{max}$, $P_{tot}$, and $\eta$ for various values of $\alpha$ with $a=0.3$, $\beta=0.03$, and $\ell=1$.}
\setlength{\tabcolsep}{-.2mm}
\begin{tabular}{|c|c|c|c|c|c|c|}
\hline
Perturbation & $\alpha$ & 0 & 0.2 & 0.4 & 0.6 & 0.8  \\
\hline
\multirow{4}{*}{Scalar}& $ \omega _{max }$ & 0.155487 & 0.172809 & 0.172809 & 0.211243 & 0.225449 \\
& $ P_{max }$ & $3.3484 \times 10^{-8}$ & $3.5996 \times 10^{-8}$& $3.8016 \times 10^{-8}$ & $4.1573\times 10^{-8}$ & $4.6306 \times 10^{-8}$\\
& $ P_{\text{tot}}$ & 3.6799$\times 10^{-9}$ & $4.2530 \times 10^{-9}$ & 4.971$\times 10^{-9}$ & 5.8881$\times 10^{-9}$ & 7.0841$\times 10^{-9}$ \\
& $ \eta $ & $1.0456\times 10^6$ & 1.1175$\times 10^6$ & 956056. & 1.2061$\times 10^6$ & 1.1419$\times 10^6$ \\
\hline
\multirow{4}{*}{Electromagnetic}& $ \omega _{max }$ & 0.14346 & 0.158634 & 0.172809 & 0.188692 & 0.20482 \\
 & $ P_{max }$ & 7.3153$\times 10^{-8}$ & 7.9163$\times 10^{-8}$ & 8.5592$\times 10^{-8}$ & 9.3133$\times 10^{-8}$ & 1.0217$\times 10^{-7}$ \\
& $ P_{\text{tot}}$ & 7.8572$\times 10^{-9}$ & 9.0809$\times 10^{-9}$ & 1.0614$\times 10^{-8}$ & 1.2572$\times 10^{-8}$ & 1.5125$\times 10^{-8}$ \\
& $ \eta$ & 416882. & 441041. & 447765. & 450731. & 441413. \\
\hline
\multirow{4}{*}{Fermionic}& $ \omega _{max }$ & 0.149055 & 0.169667 & 0.172809 & 0.211243 & 0.216399 \\
& $ P_{max }$ & 4.13349$\times 10^{-8}$ & 4.48693$\times 10^{-8}$ & 4.79750$\times 10^{-8}$ & 5.10652$\times 10^{-8}$ & 5.79986$\times 10^{-8}$ \\
& $ P_{\text{tot}}$ & 4.55163$\times 10^{-9}$ & 5.26053$\times 10^{-9}$ & 6.14893$\times 10^{-9}$ & 7.28297$\times 10^{-9}$ & 8.7623$\times 10^{-9}$ \\
& $ \eta$ & 776871. & 870933. & 772953. & 975161. & 850574. \\
\hline
\end{tabular}
\label{alpha}
\end{table}
\begin{table}[!htp]
\centering
\caption{Numerical values of $\omega_{max}$, $P_{max}$, $P_{tot}$, and $\eta$ for various values of $\beta$ with $\alpha=\beta=0.3$ and $\ell=1$.}
\setlength{\tabcolsep}{-.2mm}
\begin{tabular}{|c|c|c|c|c|c|c|}
\hline
Perturbation & $\beta$ & 0 & 0.02 & 0.04 & 0.06 & 0.08  \\
\hline
\multirow{4}{*}{Scalar}& $ \omega _{max }$ & 0.172809 & 0.172809 & 0.172809 & 0.172809 & 0.172809 \\
& $ P_{max }$ & 3.74287$\times 10^{-8}$ & 3.73788$\times 10^{-8}$ & 3.73271$\times 10^{-8}$ & 3.72738$\times 10^{-8}$ & 3.72187$\times 10^{-8}$ \\
& $ P_{\text{tot}}$ & 4.61815$\times 10^{-9}$ & 4.60014$\times 10^{-9}$ & 4.58224$\times 10^{-9}$ & 4.56444$\times 10^{-9}$ & 4.54675$\times 10^{-9}$ \\
& $ \eta$ & 1.02917$\times 10^6$ & 1.03319$\times 10^6$ & 1.03723$\times 10^6$ & 1.04127$\times 10^6$ & 1.04533$\times 10^6$ \\
\hline
\multirow{4}{*}{Electromagnetic}& $ \omega _{max }$ & 0.168486 & 0.168097 & 0.167704 & 0.167306 & 0.166905 \\
& $ P_{max }$ & 8.23909$\times 10^{-8}$ & 8.22355$\times 10^{-8}$ & 8.20810$\times 10^{-8}$ & 8.19275$\times 10^{-8}$ & 8.17748$\times 10^{-8}$ \\
& $ P_{\text{tot}}$ & 9.86054$\times 10^{-9}$ & 9.82209$\times 10^{-9}$ & 9.78387$\times 10^{-9}$ & 9.74587$\times 10^{-9}$ & 9.70809$\times 10^{-9}$ \\
& $ \eta$ & 458189. & 457861. & 457503. & 457113. & 456692. \\
\hline
\multirow{4}{*}{Fermionic}& $ \omega _{max }$ & 0.172809 & 0.172809 & 0.172809 & 0.172809 & 0.172809 \\
& $ P_{max }$ & 4.68175$\times 10^{-8}$ & 4.67335$\times 10^{-8}$ & 4.66475$\times 10^{-8}$ & 4.65595$\times 10^{-8}$ & 4.64694$\times 10^{-8}$ \\
& $ P_{\text{tot}}$ & 5.71213$\times 10^{-9}$ & 5.68985$\times 10^{-9}$ & 5.66771$\times 10^{-9}$ & 5.6457$\times 10^{-9}$ & 5.62381$\times 10^{-9}$ \\
& $ \eta$ & 832061. & 835318. & 838581. & 841851. & 845127. \\
\hline
\end{tabular}
\label{beta}
\end{table}
A significant impact of string parameter a and GUP parameter $\alpha$ on the radiation flow can be observed. The impact of $\beta$ is found to be marginal. Increasing $a$ or $\beta$ increases discontinuity in radiation flow. The sparsity, however, fluctuates with $\alpha$.

\section{Conclusion}

In this study, we thoroughly investigated GUP-corrected black holes with a string cloud. Some astrophysical phenomena, such as shadows, deflection angles, QNMs, and Hawking radiation sparsity, provide good opportunities to investigate the effects of any General Relativity alteration. We investigated the behavior of the shadows in the presence of a string cloud using three different GUP frameworks: LQGUP, QGUP, and linear GUP.  Our findings show that for all GUP corrections, increasing the value of $a$ increases both the photon sphere and the shadow radius. \\ 
Being followed by the GUP modified Shadow analysis we also studied the deflection angle in the weak form around the GUP corrected black hole with the string cloud field with accurate analysis and brought out the combined effects of string cloud $a$ and the dimensionless quantum correction parameters $\alpha$ and $\beta$ on the light rays around the black hole. Additionally, we gave details on the physical interpretation of aforementioned black hole parameters acting as attractive/repulsive gravitational charges that can physically depict to strengthen/weaken black hole gravity, thus exhibiting their distinctive futures and effects on the optical properties of GUP-corrected black holes.

Next, we looked into the quasinormal spectrum of scalar, electromagnetic, and fermionic perturbations to a GUP-corrected Schwarzschild black hole with cloud strings. For this quantum-corrected black hole, we determined the effective potential induced by the three perturbations before computing the frequencies of the QNMs using the sixth-order WKB approach. Our findings show that as the cloud string parameter $a$ and the quadratic GUP parameter $\beta$ are increased under the three field perturbations, the real part of the oscillating frequency decreases, while the negative value of the imaginary part of the oscillation frequency increases. This shows that the damping rate, or decay rate, of gravitational waves is slow. However, the impact of the linear GUP parameter $\alpha$ is the opposite.

The imprints of the underlying spacetime can also be found in the Hawking radiation. This makes it imperative to investigate various aspects of radiation to gauge the extent of influence model parameters have on Hawking radiation. To this end, we studied the qualitative as well as quantitative aspects of the power spectrum and sparsity. Our inquisition revealed that the power emitted decreases with an increase in the string parameter $a$ and the GUP parameter $\beta$, although the variation with $\beta$ is marginal. On the other hand, the emitted power increases with the GUP parameter $\alpha$. It is also observed that the frequency where $P_{\ell}$ peaks decreases with $a$ and $\beta$ but increases with $\alpha$. We next elucidate the impact of model parameters on the continuity of radiation flow through a dimensionless parameter $\eta$ signifying sparsity. The values of $\eta$ provided in Tables \ref{a}, \ref{alpha} and  \ref{beta} are much greater than unity, implying a discontinuous flow of radiation. The sparsity of radiation flow increases with $a$ or $\beta$ but fluctuates with $\alpha$. The impact of the string parameter a and GUP parameter $\alpha$ is significant. However, the sparsity varies marginally with $\beta$. These observations show that astrophysical observations concerning Hawking radiation can help us validate our model in question. 
\\ \begin{acknowledgments}
We are thankful to the Editor and anonymous Referees for their constructive suggestions and comments. This work is supported by the National Natural Science Foundation of China under Grant No. 11675143 and the National Key Research and Development Program of China under Grant No. 2020YFC2201503. 
\end{acknowledgments}

\end{document}